\documentclass[twoside]{dis09}
\usepackage[latin1]{inputenc}
\usepackage[dvips]{graphicx,epsfig,color}
\usepackage{wrapfig,rotating}
\usepackage{amssymb,amsmath,array}

\begin{document}

%%%%%%%%%%%%%%%%%%%%%%%%%%%%%

\newcommand{\invfb}{\ensuremath{\mathrm{fb^{-1}}}}
\newcommand{\invpb}{\ensuremath{\mathrm{pb^{-1}}}}
\newcommand{\invnb}{\ensuremath{\mathrm{nb^{-1}}}}
\newcommand{\abt} {\mbox{$|t|$}}
\newcommand{\rhoprim}{\mbox{$\rho^\prime$}}
\newcommand{\tprim}{\mbox{$t^\prime$}}
\newcommand{\rhop}{\mbox{$\rho^\prime$}}
\newcommand{\zet}{\mbox{$\zeta$}}
\newcommand{\rfivecomb}{\mbox{$r^5_{00} + 2 r^5_{11}$}}
\newcommand{\ronecomb}{\mbox{$r^1_{00} + 2 r^1_{11}$}}
\newcommand{\gstarVM} {\mbox{$\gamma^* \ \!p \rightarrow V\ \!Y$}}
\newcommand{\gstarVMel} {\mbox{$\gamma^* \ \!p \rightarrow V\ \!p$}}
\newcommand{\gsrel} {\mbox{$\gamma^* \ \!p \rightarrow \rho \ \!p$}}
\newcommand{\gsrpd} {\mbox{$\gamma^* \ \!p \rightarrow \rho \ \!Y$}}
\newcommand{\gspel} {\mbox{$\gamma^* \ \!p \rightarrow \phi \ \!p$}}
\newcommand{\gsppd} {\mbox{$\gamma^* \ \!p \rightarrow \phi \ \!Y$}}
\newcommand{\gstarp} {\mbox{$\gamma^*\ \!p$}}
\newcommand{\mv} {\mbox{$M_V$}}
\newcommand{\mvsq} {\mbox{$M_V^2$}}
\newcommand{\msq} {\mbox{$M_V^2$}}
\newcommand{\qsqplmsq} {\mbox{($Q^2 \!+ \!M_V^2$})}
\newcommand{\qqsqplmsq} {\mbox{$Q^2 \!+ \!M_V^2}$}
\newcommand{\alprim}{\mbox{$\alpha^\prime$}}
\newcommand{\alphaz}{\mbox{$\alpha(0)$}}
\newcommand{\alpomz}{\mbox{$\alpha_{\PO}(0)$}}
\newcommand{\hence}{\mbox{$=>$}}
\newcommand{\vm}{\mbox{$V\!M$}}
\newcommand{\sur}{\mbox{\ \! / \ \!}}
\newcommand{\tzz} {\mbox{$T_{00}$}}
\newcommand{\tuu} {\mbox{$T_{11}$}}
\newcommand{\tzu} {\mbox{$T_{01}$}}
\newcommand{\tuz} {\mbox{$T_{10}$}}
\newcommand{\tmuu} {\mbox{$T_{-11}$}}
\newcommand{\tumu} {\mbox{$T_{1-1}$}}
\newcommand{\ralpha} {\mbox{$\tuu / \tzz$}}
\newcommand{\rbeta} {\mbox{$\tzu / \tzz$}}
\newcommand{\rdelta} {\mbox{$\tuz / \tzz$}}
\newcommand{\reta} {\mbox{$\tmuu / \tzz$}}
\newcommand{\abstzz} {\mbox{$|T_{00}|$}}
\newcommand{\abstuu} {\mbox{$|T_{11}|$}}
\newcommand{\abstzu} {\mbox{$|T_{01}|$}}
\newcommand{\abstuz} {\mbox{$|T_{10}|$}}
\newcommand{\abstmuu} {\mbox{$|T_{-11}|$}}
\newcommand{\rralpha} {\mbox{\abstuu \sur \abstzz}}
\newcommand{\rrbeta} {\mbox{\abstzu \sur \abstzz}}
\newcommand{\rrdelta} {\mbox{\tuz \sur \tzz}}
\newcommand{\rreta} {\mbox{\abstmuu \sur \abstzz}}
\newcommand{\averm} {\mbox{$\av {M}$}}
\newcommand{\rapproch} {\mbox{$R_{SCHC+T_{01}}$}}
\newcommand{\chisq} {\mbox{$\chi^2 / {\rm d.o.f.}$}}

\newcommand{\scaleqsqplmsq} {\mbox{$\qsqplmsq /4$}}

%%%%%%%%%%%%%%%%%%%%%%%%%%%%%

%%%%%%%%%%%%%%%%%%%%%%%%%%%%%%%%%%%%%%%%%%%%%%%%%%%%%%%%%%%%%%%%%%%%%%
%                                                                    %
%                                                                    %
%           DEFINITIONS OF NEW COMMANDS IN LATEX                     %
%                                                                    %
%                                                                    %
%%%%%%%%%%%%%%%%%%%%%%%%%%%%%%%%%%%%%%%%%%%%%%%%%%%%%%%%%%%%%%%%%%%%%%
%
%=====================================================================
% kinematics
%
\newcommand{\s}{\mbox{$s$}}
\newcommand{\ttra}{\mbox{$t$}}
\newcommand{\modt}{\mbox{$|t|$}}
\newcommand{\eminpz}{\mbox{$E-p_z$}}
\newcommand{\eminpzs}{\mbox{$\Sigma(E-p_z)$}}
\newcommand{\rap}{\ensuremath{\eta^*} }
\newcommand{\W}{\mbox{$W$}}
\newcommand{\w}{\mbox{$W$}}
\newcommand{\Q}{\mbox{$Q$}}
\newcommand{\q}{\mbox{$Q$}}
\newcommand{\xB}{\mbox{$x$}}  % Bjorken x
\newcommand{\xF}{\mbox{$x_F$}}  % Feynman x
\newcommand{\xg}{\mbox{$x_g$}}  % x_g
\newcommand{\xbj}{x}
\newcommand{\xpom}{x_{\PO}}
\newcommand{\y}{\mbox{$y~$}}
\newcommand{\Qsq}{\mbox{$Q^2$}}
\newcommand{\qsq}{\mbox{$Q^2$}}
\newcommand{\kjet}{\mbox{$k_{T\rm{jet}}$}}
\newcommand{\xjet}{\mbox{$x_{\rm{jet}}$}}
\newcommand{\Ejet}{\mbox{$E_{\rm{jet}}$}}
\newcommand{\thjet}{\mbox{$\theta_{\rm{jet}}$}}
\newcommand{\pjet}{\mbox{$p_{T\rm{jet}}$}}
\newcommand{\et}{\mbox{$E_T~$}}
\newcommand{\kt}{\mbox{$k_T~$}}
\newcommand{\ptrans}{\mbox{$p_T~$}}
\newcommand{\pth}{\mbox{$p_T^h~$}}
\newcommand{\pte}{\mbox{$p_T^e~$}}
\newcommand{\ptsq}{\mbox{$p_T^{\star 2}~$}}
\newcommand{\as}{\mbox{$\alpha_s~$}}
\newcommand{\ycut}{\mbox{$y_{\rm cut}~$}}
\newcommand{\gx}{\mbox{$g(x_g,Q^2)$~}}
\newcommand{\xpart}{\mbox{$x_{\rm part~}$}}
\newcommand{\mrsdm}{\mbox{${\rm MRSD}^-~$}}
\newcommand{\mrsdmp}{\mbox{${\rm MRSD}^{-'}~$}}
\newcommand{\mrsdn}{\mbox{${\rm MRSD}^0~$}}
\newcommand{\lambdams}{\mbox{$\Lambda_{\rm \bar{MS}}~$}}
%
%=====================================================================
% section efficace abbreviations
%
\newcommand{\gp}{\ensuremath{\gamma}p }
\newcommand{\gammasp}{\ensuremath{\gamma}*p }
\newcommand{\gammap}{\ensuremath{\gamma}p }
\newcommand{\gsp}{\ensuremath{\gamma^*}p }
\newcommand{\dsiget}{\ensuremath{{\rm d}\sigma_{ep}/{\rm d}E_t^*} }
\newcommand{\dsigrap}{\ensuremath{{\rm d}\sigma_{ep}/{\rm d}\eta^*} }
% ep 
\newcommand{\epem}{\mbox{$e^+e^-$}}
\newcommand{\ep}{\mbox{$ep~$}}
\newcommand{\epl}{\mbox{$e^{+}$}}
\newcommand{\emi}{\mbox{$e^{-}$}}
\newcommand{\epm}{\mbox{$e^{\pm}$}}
\newcommand{\se}{section efficace}
\newcommand{\ses}{sections efficaces}
%
%=====================================================================
% elastic VM abbreviations 
%
% VM
\newcommand{\phib}{\mbox{$\varphi$}}
\newcommand{\rh}{\mbox{$\rho$}}
\newcommand{\rhz}{\mbox{$\rh^0$}}
\newcommand{\ph}{\mbox{$\phi$}}
\newcommand{\om}{\mbox{$\omega$}}
\newcommand{\jpsi}{\mbox{$J/\psi$}}
\newcommand{\pipi}{\mbox{$\pi^+\pi^-$}}
\newcommand{\pip}{\mbox{$\pi^+$}}
\newcommand{\pim}{\mbox{$\pi^-$}}
\newcommand{\kk}{\mbox{K^+K^-$}}
% b parameter
\newcommand{\bsl}{\mbox{$b$}}
\newcommand{\alp}{\mbox{$\alpha^\prime$}}
\newcommand{\alpom}{\mbox{$\alpha_{\PO}$}}
\newcommand{\alpomp}{\mbox{$\alpha_{\PO}^\prime$}}
% polarisation
\newcommand{\rzzzz}{\mbox{$r_{00}^{04}$}}
\newcommand{\rzqzz}{\mbox{$r_{00}^{04}$}}
\newcommand{\rzquz}{\mbox{$r_{10}^{04}$}}
\newcommand{\rzqumu}{\mbox{$r_{1-1}^{04}$}}
\newcommand{\ruuu}{\mbox{$r_{11}^{1}$}}
\newcommand{\ruzz}{\mbox{$r_{00}^{1}$}}
\newcommand{\ruuz}{\mbox{$r_{10}^{1}$}}
\newcommand{\ruumu}{\mbox{$r_{1-1}^{1}$}}
\newcommand{\rduz}{\mbox{$r_{10}^{2}$}}
\newcommand{\rdumu}{\mbox{$r_{1-1}^{2}$}}
\newcommand{\rcuu}{\mbox{$r_{11}^{5}$}}
\newcommand{\rczz}{\mbox{$r_{00}^{5}$}}
\newcommand{\rcuz}{\mbox{$r_{10}^{5}$}}
\newcommand{\rcumu}{\mbox{$r_{1-1}^{5}$}}
\newcommand{\rsuz}{\mbox{$r_{10}^{6}$}}
\newcommand{\rsumu}{\mbox{$r_{1-1}^{6}$}}
\newcommand{\rzqik}{\mbox{$r_{ik}^{04}$}}
\newcommand{\rhzik}{\mbox{$\rh_{ik}^{0}$}}
\newcommand{\rhqik}{\mbox{$\rh_{ik}^{4}$}}
\newcommand{\rhaik}{\mbox{$\rh_{ik}^{\alpha}$}}
\newcommand{\rhzzz}{\mbox{$\rh_{00}^{0}$}}
\newcommand{\rhqzz}{\mbox{$\rh_{00}^{4}$}}
\newcommand{\raik}{\mbox{$r_{ik}^{\alpha}$}}
\newcommand{\razz}{\mbox{$r_{00}^{\alpha}$}}
\newcommand{\rauz}{\mbox{$r_{10}^{\alpha}$}}
\newcommand{\raumu}{\mbox{$r_{1-1}^{\alpha}$}}

\newcommand{\R}{\mbox{$R$}}
\newcommand{\rzero}{\mbox{$r_{00}^{04}$}}
\newcommand{\rone}{\mbox{$r_{1-1}^{1}$}}
\newcommand{\costh}{\mbox{$\cos\theta$}}
\newcommand{\cosp}{\mbox{$\cos\psi$}}
\newcommand{\costop}{\mbox{$\cos(2\psi)$}}
\newcommand{\cosd}{\mbox{$\cos\delta$}}
\newcommand{\cossqp}{\mbox{$\cos^2\psi$}}
\newcommand{\cossqt}{\mbox{$\cos^2\theta^*$}}
\newcommand{\sint}{\mbox{$\sin\theta^*$}}
\newcommand{\sintot}{\mbox{$\sin(2\theta^*)$}}
\newcommand{\sinsqt}{\mbox{$\sin^2\theta^*$}}
\newcommand{\costhst}{\mbox{$\cos\theta^*$}}
\newcommand{\vep}{\mbox{$V p$}}
% mass
\newcommand{\mpipi}{\mbox{$m_{\pi^+\pi^-}$}}
\newcommand{\mkk}{\mbox{$m_{KK}$}}
\newcommand{\mkaka}{\mbox{$m_{K^+K^-}$}}
\newcommand{\mpp}{\mbox{$m_{\pi\pi}$}}       %  for use in B_W
\newcommand{\mppsq}{\mbox{$m_{\pi\pi}^2$}}   %  for use in B_W
\newcommand{\mpi}{\mbox{$m_{\pi}$}}          %  for use in B_W
\newcommand{\mrho}{\mbox{$m_{\rho}$}}        %  for use in B_W
\newcommand{\mrhosq}{\mbox{$m_{\rho}^2$}}    %  for use in B_W
% width
\newcommand{\Gmpp}{\mbox{$\Gamma (\mpp)$}}   %  for use in B_W
\newcommand{\Gmppsq}{\mbox{$\Gamma^2(\mpp)$}}%  for use in B_W
\newcommand{\Grho}{\mbox{$\Gamma_{\rho}$}}   %  for use in B_W
\newcommand{\grho}{\mbox{$\Gamma_{\rho}$}}   %  for use in B_W
\newcommand{\Grhosq}{\mbox{$\Gamma_{\rho}^2$}}   %  for use in B_W
%
%=====================================================================
% units
%
\newcommand{\cm}{\mbox{\rm cm}}
\newcommand{\GeV}{\mbox{\rm GeV}}
\newcommand{\gev}{\mbox{\rm GeV}}
\newcommand{\GeVx}{\rm GeV}
\newcommand{\gevx}{\rm GeV}
\newcommand{\GeVc}{\rm GeV/c}
\newcommand{\gevc}{\rm GeV/c}
\newcommand{\MeVc}{\rm MeV/c}
\newcommand{\mevc}{\rm MeV/c}
\newcommand{\MeV}{\mbox{\rm MeV}}
\newcommand{\mev}{\mbox{\rm MeV}}
\newcommand{\MeVx}{\mbox{\rm MeV}}
\newcommand{\mevx}{\mbox{\rm MeV}}
\newcommand{\GeVsq}{\mbox{${\rm GeV}^2$}}
\newcommand{\gevsq}{\mbox{${\rm GeV}^2$}}
\newcommand{\gevsqc}{\mbox{${\rm GeV^2/c^4}$}}
\newcommand{\gevcsq}{\mbox{${\rm GeV/c^2}$}}
\newcommand{\mevcsq}{\mbox{${\rm MeV/c^2}$}}
\newcommand{\GeVsqm}{\mbox{${\rm GeV}^{-2}$}}
\newcommand{\gevsqm}{\mbox{${\rm GeV}^{-2}$}}
\newcommand{\nb}{\mbox{${\rm nb}$}}
\newcommand{\nbinv}{\mbox{${\rm nb^{-1}}$}}
\newcommand{\pbinv}{\mbox{${\rm pb^{-1}}$}}
\newcommand{\mm}{\mbox{$\cdot 10^{-2}$}}
\newcommand{\mmm}{\mbox{$\cdot 10^{-3}$}}
\newcommand{\mmmm}{\mbox{$\cdot 10^{-4}$}}
\newcommand{\degr}{\mbox{$^{\circ}$}}
%
%=====================================================================
% F2
%
\newcommand{\F}{$ F_{2}(x,Q^2)\,$}  
\newcommand{\Fc}{$ F_{2}\,$}    
\newcommand{\XP}{x_{{I\!\!P}/{p}}}       
\newcommand{\TOSS}{x_{{i}/{\PO}}}        
\newcommand{\un}[1]{\mbox{\rm #1}} 
\newcommand{\LO}{Leading Order}
\newcommand{\NLO}{Next to Leading Order}
\newcommand{\ft}{$ F_{2}\,$}
%
%=====================================================================
%  latex command abbreviations 
%
\newcommand{\mc}{\multicolumn}
\newcommand{\bce}{\begin{center}}
\newcommand{\ece}{\end{center}}
\newcommand{\beq}{\begin{equation}}
\newcommand{\eeq}{\end{equation}}
\newcommand{\bea}{\begin{eqnarray}}
\newcommand{\eea}{\end{eqnarray}}
%
%=====================================================================
% inequation symbols
%
%less than or approx. symbol
\def\lsim{\mathrel{\rlap{\lower4pt\hbox{\hskip1pt$\sim$}}
    \raise1pt\hbox{$<$}}}         
%greater than or approx. symbol
\def\gsim{\mathrel{\rlap{\lower4pt\hbox{\hskip1pt$\sim$}}
    \raise1pt\hbox{$>$}}}         
%
%=====================================================================
% diffrative abbreviations : F2d3
%
\newcommand{\pom}{{I\!\!P}}
\newcommand{\PO}{I\!\!P}
\newcommand{\slowpi}{\pi_{\mathit{slow}}}
\newcommand{\fiidiii}{F_2^{D(3)}}
\newcommand{\fiidiiiarg}{\fiidiii\,(\beta,\,Q^2,\,x)}
\newcommand{\n}{1.19\pm 0.06 (stat.) \pm0.07 (syst.)}
\newcommand{\nz}{1.30\pm 0.08 (stat.)^{+0.08}_{-0.14} (syst.)}
\newcommand{\fiidiiiful}{F_2^{D(4)}\,(\beta,\,Q^2,\,x,\,t)}
\newcommand{\fiipom}{\tilde F_2^D}
\newcommand{\ALPHA}{1.10\pm0.03 (stat.) \pm0.04 (syst.)}
\newcommand{\ALPHAZ}{1.15\pm0.04 (stat.)^{+0.04}_{-0.07} (syst.)}
\newcommand{\fiipomarg}{\fiipom\,(\beta,\,Q^2)}
\newcommand{\pomflux}{f_{\pom / p}}
\newcommand{\nxpom}{1.19\pm 0.06 (stat.) \pm0.07 (syst.)}
\newcommand {\gapprox}
   {\raisebox{-0.7ex}{$\stackrel {\textstyle>}{\sim}$}}
\newcommand {\lapprox}
   {\raisebox{-0.7ex}{$\stackrel {\textstyle<}{\sim}$}}
\newcommand{\pomfluxarg}{f_{\pom / p}\,(x_\pom)}
\newcommand{\dsf}{\mbox{$F_2^{D(3)}$}}
\newcommand{\dsfva}{\mbox{$F_2^{D(3)}(\beta,Q^2,x_{I\!\!P})$}}
\newcommand{\dsfvb}{\mbox{$F_2^{D(3)}(\beta,Q^2,x)$}}
\newcommand{\dsfpom}{$F_2^{I\!\!P}$}
\newcommand{\gap}{\stackrel{>}{\sim}}
\newcommand{\lap}{\stackrel{<}{\sim}}
\newcommand{\fem}{$F_2^{em}$}
\newcommand{\tsnmp}{$\tilde{\sigma}_{NC}(e^{\mp})$}
\newcommand{\tsnm}{$\tilde{\sigma}_{NC}(e^-)$}
\newcommand{\tsnp}{$\tilde{\sigma}_{NC}(e^+)$}
\newcommand{\st}{$\star$}
\newcommand{\sst}{$\star \star$}
\newcommand{\ssst}{$\star \star \star$}
\newcommand{\sssst}{$\star \star \star \star$}
\newcommand{\tw}{\theta_W}
\newcommand{\sw}{\sin{\theta_W}}
\newcommand{\cw}{\cos{\theta_W}}
\newcommand{\sww}{\sin^2{\theta_W}}
\newcommand{\cww}{\cos^2{\theta_W}}
\newcommand{\trm}{m_{\perp}}
\newcommand{\trp}{p_{\perp}}
\newcommand{\trmm}{m_{\perp}^2}
\newcommand{\trpp}{p_{\perp}^2}
\newcommand{\ev}{\'ev\'enement}
\newcommand{\evs}{\'ev\'enements}
\newcommand{\mdv}{mod\`ele \`a dominance m\'esovectorielle}
\newcommand{\mdmv}{mod\`ele \`a dominance m\'esovectorielle}
\newcommand{\mdm}{mod\`ele \`a dominance m\'esovectorielle}
\newcommand{\idiff}{interaction diffractive}
\newcommand{\idiffs}{interactions diffractives}
\newcommand{\pdmv}{production diffractive de m\'esons vecteurs}
\newcommand{\pdmr}{production diffractive de m\'esons \rh}
\newcommand{\pdmp}{production diffractive de m\'esons \ph}
\newcommand{\pdmo}{production diffractive de m\'esons \om}
\newcommand{\pdm}{production diffractive de m\'esons}
\newcommand{\pdiff}{production diffractive}
\newcommand{\diff}{diffractive}
\newcommand{\produ}{production}
\newcommand{\mvs}{m\'esons vecteurs}
\newcommand{\me}{m\'eson}
\newcommand{\mr}{m\'eson \rh}
\newcommand{\mph}{m\'eson \ph}
\newcommand{\mo}{m\'eson \om}
\newcommand{\mrs}{m\'esons \rh}
\newcommand{\mps}{m\'esons \ph}
\newcommand{\mos}{m\'esons \om}
\newcommand{\photo}{photoproduction}
\newcommand{\agq}{\`a grand \qsq}
\newcommand{\agqsq}{\`a grand \qsq}
\newcommand{\apq}{\`a petit \qsq}
\newcommand{\apqsq}{\`a petit \qsq}
\newcommand{\de}{d\'etecteur}
%
%=====================================================================
% alpha s
%
\newcommand{\sqrts}{$\sqrt{s}$}
\newcommand{\Oa}{$O(\alpha_s)$}
\newcommand{\Oaa}{$O(\alpha_s^2)$}
\newcommand{\PT}{p_{\perp}}
\newcommand{\sh}{\hat{s}}
\newcommand{\uh}{\hat{u}}
\newcommand{\ttbs}{\char'134}
\newcommand{\xpomlo}{3\times10^{-4}}
\newcommand{\xpomup}{0.05}
\newcommand{\llq}{$\alpha_s \ln{(\qsq / \Lambda_{QCD}^2)}$}
\newcommand{\llqx}{$\alpha_s \ln{(\qsq / \Lambda_{QCD}^2)} \ln{(1/x)}$}
\newcommand{\llx}{$\alpha_s \ln{(1/x)}$}
%
%=====================================================================
% name groups
%
\newcommand{\Brodsky}{Brodsky {\it et al.}}
\newcommand{\FKS}{Frankfurt, Koepf and Strikman}
\newcommand{\Kop}{Kopeliovich {\it et al.}}
\newcommand{\Ginzburg}{Ginzburg {\it et al.}}
\newcommand{\Ryskin}{\mbox{Ryskin}}
\newcommand{\Kaidalov}{Kaidalov {\it et al.}}
%
%=====================================================================
% journals
%
\def\ar#1#2#3   {{\em Ann. Rev. Nucl. Part. Sci.} {\bf#1} (#2) #3}
\def\epj#1#2#3  {{\em Eur. Phys. J.} {\bf#1} (#2) #3}
\def\err#1#2#3  {{\it Erratum} {\bf#1} (#2) #3}
\def\ib#1#2#3   {{\it ibid.} {\bf#1} (#2) #3}
\def\ijmp#1#2#3 {{\em Int. J. Mod. Phys.} {\bf#1} (#2) #3}
\def\jetp#1#2#3 {{\em JETP Lett.} {\bf#1} (#2) #3}
\def\mpl#1#2#3  {{\em Mod. Phys. Lett.} {\bf#1} (#2) #3}
\def\nim#1#2#3  {{\em Nucl. Instr. Meth.} {\bf#1} (#2) #3}
\def\nc#1#2#3   {{\em Nuovo Cim.} {\bf#1} (#2) #3}
\def\np#1#2#3   {{\em Nucl. Phys.} {\bf#1} (#2) #3}
\def\pl#1#2#3   {{\em Phys. Lett.} {\bf#1} (#2) #3}
\def\prep#1#2#3 {{\em Phys. Rep.} {\bf#1} (#2) #3}
\def\prev#1#2#3 {{\em Phys. Rev.} {\bf#1} (#2) #3}
\def\prl#1#2#3  {{\em Phys. Rev. Lett.} {\bf#1} (#2) #3}
\def\ptp#1#2#3  {{\em Prog. Th. Phys.} {\bf#1} (#2) #3}
\def\rmp#1#2#3  {{\em Rev. Mod. Phys.} {\bf#1} (#2) #3}
\def\rpp#1#2#3  {{\em Rep. Prog. Phys.} {\bf#1} (#2) #3}
\def\sjnp#1#2#3 {{\em Sov. J. Nucl. Phys.} {\bf#1} (#2) #3}
\def\spj#1#2#3  {{\em Sov. Phys. JEPT} {\bf#1} (#2) #3}
\def\zp#1#2#3   {{\em Zeit. Phys.} {\bf#1} (#2) #3}
%
%=====================================================================
% non classified
%
\newcommand{\clearemptydoublepage}{\newpage{\pagestyle{empty}\cleardoublepage}}
\newcommand{\scaption}[1]{\caption{\protect{\footnotesize  #1}}}
\newcommand{\proc}[2]{\mbox{$ #1 \rightarrow #2 $}}
\newcommand{\average}[1]{\mbox{$ \langle #1 \rangle $}}
\newcommand{\av}[1]{\mbox{$ \langle #1 \rangle $}}

%%%%%%%%%%%%%%%%%%%%%%%%%%%%%

%%%%%%%%%%%%%%%%%%%%%%%%%%%%%

\title{DVCS and Vector Meson Production with H1}

%***********************************************************************
% AUTHORS INFORMATION AREA
%***********************************************************************
\author{Pierre Marage
\vspace{.3cm}\\
Faculté des Sciences - Universit\'e Libre de Bruxelles \\
ULB-CP 230 - Boulevard du Triomphe - B-1050 Bruxelles - Belgium
}
%***********************************************************************
% END OF AUTHORS INFORMATION AREA
%***********************************************************************

\maketitle

\begin{abstract}
Recent results on Deeply Virtual Compton Scattering (DVCS) and 
exclusive vector meson (VM)
production obtained with the H1 detector at HERA are reviewed.
Emphasis is put on the transition from soft to hard diffraction and on spin
dynamics~\cite{url}.

\end{abstract}

%%%%%%%%%%%%%%%%%%%%%%%%%%%%%

\section {Introduction}
Studies of the DVCS process and of exclusive VM production 
at HERA provide,
in the presence of a hard scale, unique information on the mechanism of 
diffraction, in particular on the transition from soft to hard diffraction and 
spin dynamics.

Exclusive final states studied by the H1 collaboration include
real photons~\cite{dvcs,high-t-photons} and
light ($\rho$ and \ph\ ~\cite{rho-photoprod,rho-hera1,high-t-rho})
and heavy VMs 
(\jpsi~\cite{jpsi,high-t-jpsi}, 
$\psi(2s)$~\cite{Adloff:2002re} 
and $\Upsilon$~\cite{upsilon_h1}),
in the elastic and proton dissociation channels.
A hard scale is provided by the VM mass $M_V$, 
by the negative square of the photon four-momentum, \qsq\
(with $\qsq \simeq 0$ for photoproduction and 
$1.5 \leq Q^2 \leq 90~\gevsq$ for electroproduction),
or by the square of the four-momentum transfer at the proton vertex, $t$.
Cross sections are expressed in terms of $\gamma^*p$ scattering.

The data are interpreted in terms of two complementary QCD
approaches.
Following a collinear factorisation theorem, the electroproduction of 
light VMs by longitudinally polarised photons and the production of heavy VMs 
can be described by the convolution of the hard process with generalised parton 
distributions in the proton (GPDs).
High energy DVCS and VM production can also be described as the
factorisation of virtual photon fluctuation into a $q \bar q$ colour dipole, 
diffractive dipole--proton scattering, 
and $q \bar q$ recombination into the final state photon or VM.
The interaction scale $\mu$ is given by the characteristic transverse size of 
the dipole, with $\mu^2 \simeq \scaleqsqplmsq$ for \jpsi\ production and light 
VM production by longitudinal photons, 
whereas this value may be significantly reduced 
for light VM production by transversely polarised photons, 
because of end-point contributions in the photon wave function.
For DVCS, the role of LO contributions support a scale of the order of \qsq.
Several models based on either approach are compared to the H1 data; 
references and brief descriptions can be found in~\cite{dvcs,rho-hera1}.

High \modt\ photoproduction data with 
$2 \leq \modt \leq 30~\gevsq$~\cite{high-t-photons,high-t-rho,high-t-jpsi}, which 
are not further presented here,
tend to support BFKL evolution.

%%%%%%%%%%%%%%%%%%%%%%%%%%%%%
\section{Kinematic dependences}
\begin{figure}[htbp]
\begin{center}
\includegraphics[width=0.40\columnwidth]{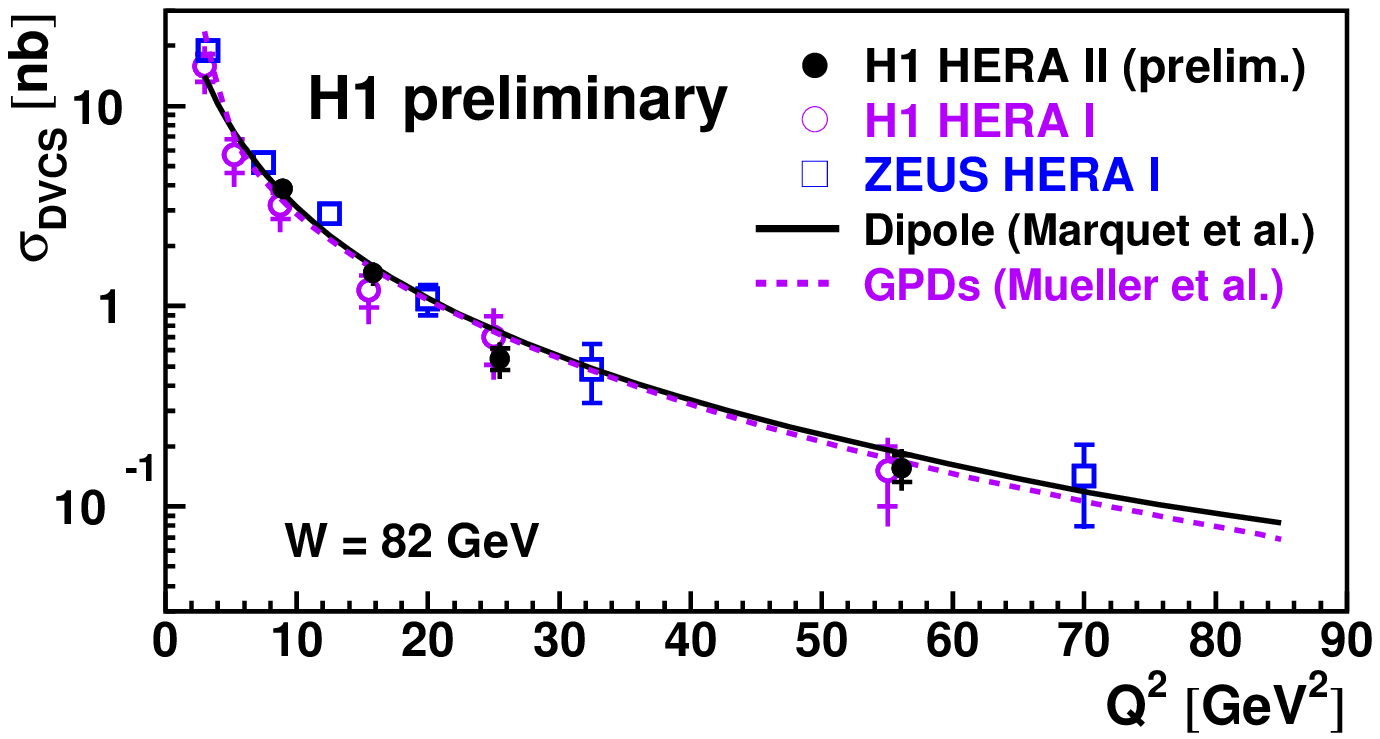}
\includegraphics[width=0.36\columnwidth]{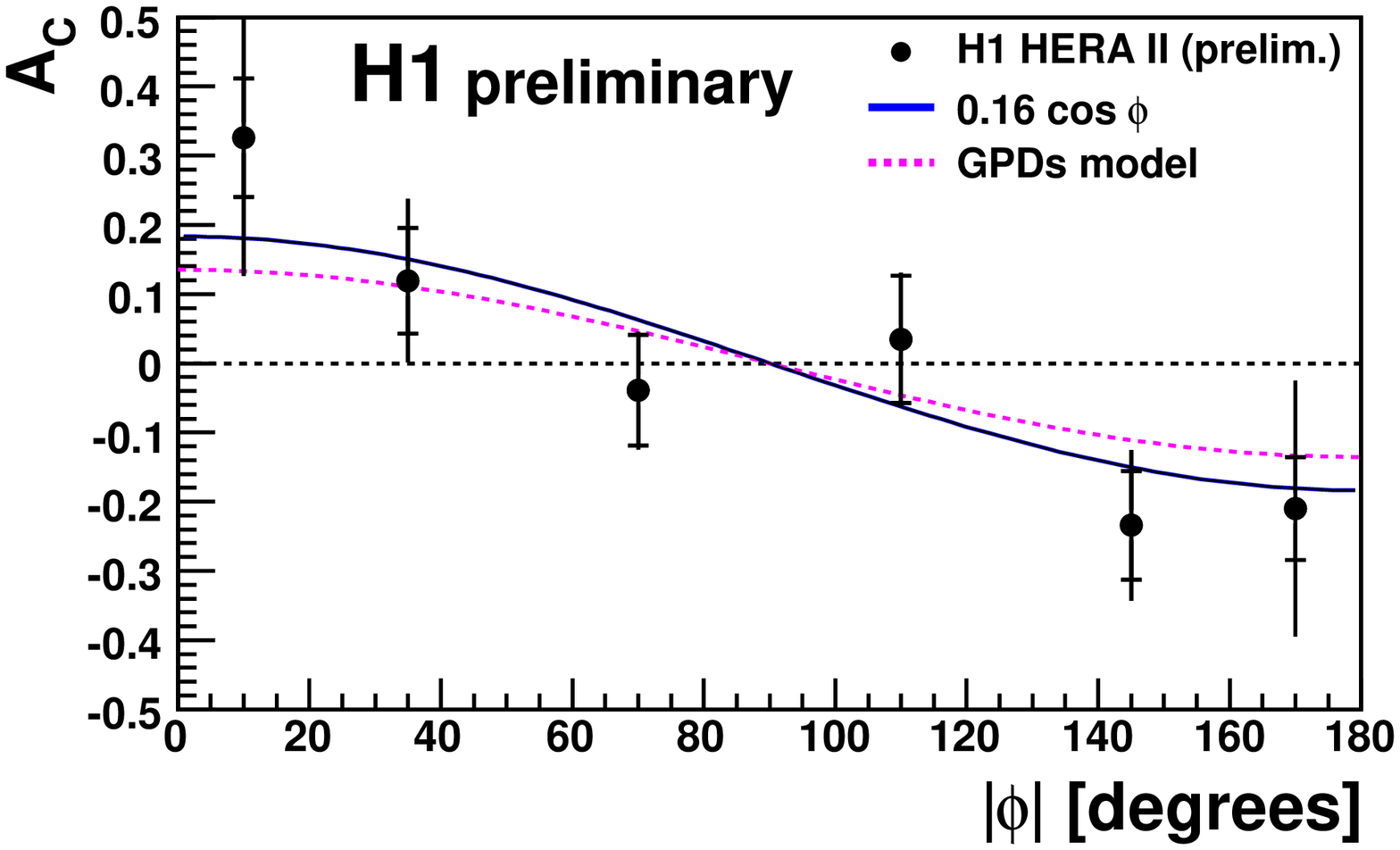}
\vspace{-0.3cm}
\caption{(left) \qsq\ dependence of DVCS production and model predictions;
(right) beam charge asymmetry, $\cos \phi$ fit and predictions of a GPD 
model~\protect\cite{dvcs}.}
\label{fig:dvcs}
\end{center}
\vspace{-0.3cm}
\end{figure}

\begin{figure}[h]
\vspace{-0.5cm}
\begin{minipage}{0.48\columnwidth}
%\vspace{-0.45cm}
\centerline{\includegraphics[width=0.80\columnwidth]{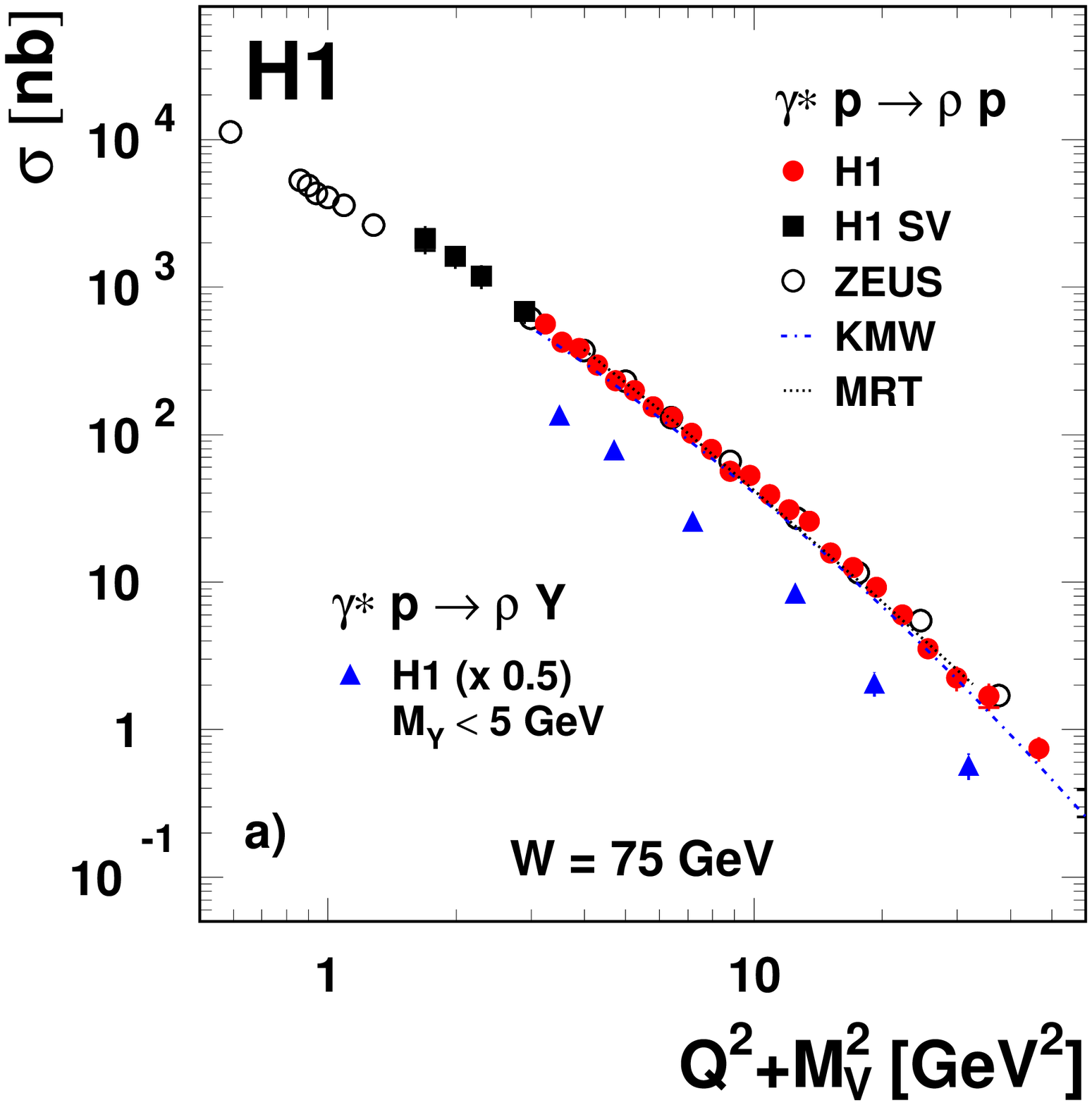}}
%\centerline{\includegraphics[width=0.80\columnwidth]{figures/hera1_rho_q2.eps}}
\vspace{-0.3cm}
\caption{\qsq\ dependence of elastic and proton dissociative electroproduction 
cross sections of \rh\ mesons, with model predictions~\protect\cite{rho-hera1}.}
\label{fig:rho_q2}
\end{minipage}
\hspace{2mm}
\begin{minipage}[h]{0.48\columnwidth}
\centerline{\includegraphics[width=0.80\columnwidth]{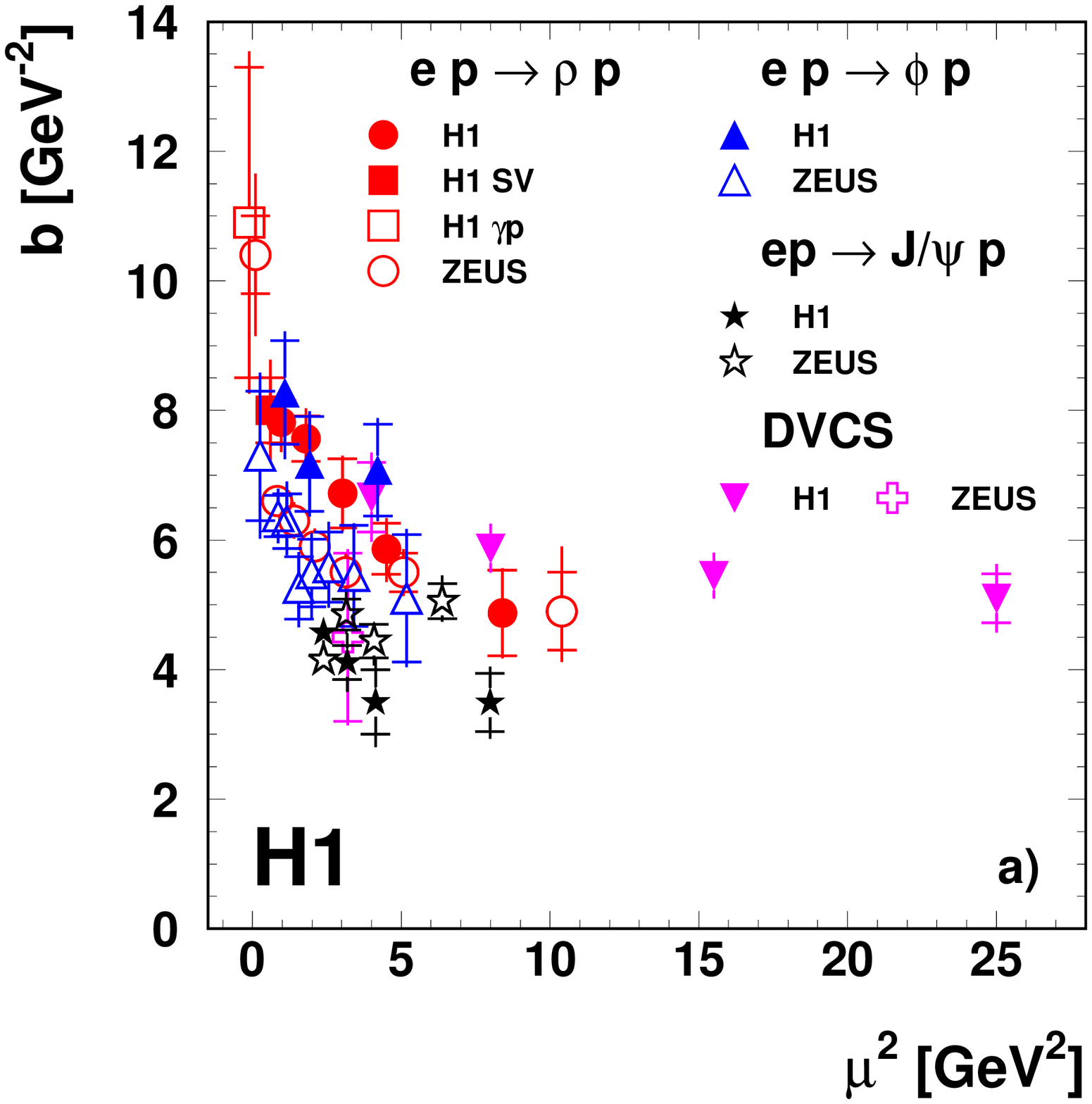}}
%\centerline{\includegraphics[width=0.80\columnwidth]{figures/hera1_vm_el_bq2.eps}}
\vspace{-0.3cm}
\caption{Elastic $b$ slopes, as a function of $\mu^2 = \scaleqsqplmsq$
for VM production and $\mu^2 = \qsq$ for DVCS.}
\label{fig:b-slopes}
\end{minipage}
\vspace{-0.3cm}
\end{figure}

%%%%%%%%%%%%%%%%%%%%%%%%%%%%%
%
%%%%%%%%%%%%%%%%%%%%%%%%%%%%%
%%%%%%%%%%%%%%%%%%%%%%%%%%%%%
\paragraph {{\bf DVCS}}
%%%%%%%%%%%%%%%%%%%%%%%%%%%%%
The \qsq\ evolution of DVCS production~\cite{dvcs}, presented in
Fig.~\ref{fig:dvcs}, is well described by models both using GPDs 
or a dipole approach.
The interference of the DVCS and Bethe-Heitler processes gives access, 
through the measurement of beam charge asymmetry, to the ratio $\rho$ of the
real to imaginary parts of the DVCS amplitude.
The measurement is $\rho = 0.20 \pm 0.05 \pm 0.08$, in agreement with 
the value $\rho = 0.25 \pm 0.03 \pm 0.05$ obtained from a dispersion relation 
using the $W$ dependence of the cross section.
%
%--------------------------------------------------------------------------------------------------------
%--------------------------------------------------------------------------------------------------------
%
%%%%%%%%%%%%%%%%%%%%%%%%%%%%%
%%%%%%%%%%%%%%%%%%%%%%%%%%%%%
\paragraph {{\bf Light VM electroproduction}}
%%%%%%%%%%%%%%%%%%%%%%%%%%%%%
The cross sections for elastic and proton dissociative \rh\ and \ph\ 
electroproduction have been measured with high precision~\cite{rho-hera1}.
The \qsq\ dependence, shown for \rh\ mesons in Fig.~\ref{fig:rho_q2}, is reasonably 
described by several models, which predict separately the longitudinal and 
transverse cross sections (for polarised cross sections, see~\cite{rho-hera1}).
%--------------------------------------------------------------------------------------------------------
%--------------------------------------------------------------------------------------------------------
%%%%%%%%%%%%%%%%%%%%%%%%%%%%%
%%%%%%%%%%%%%%%%%%%%%%%%%%%%%
\paragraph {{\bf $t$ slopes}}
%%%%%%%%%%%%%%%%%%%%%%%%%%%%%
The \modt\ distributions for DVCS, light and heavy VM production, both in the 
elastic channel with $\modt\ \leq 0.5~\gevsq$ and in the proton dissociative 
channel with $\modt\ \lapprox\ 2.5~\gevsq$, are exponentially falling, with
$\rm {d} \sigma / \rm {d}t \propto e^{-b |t|}$.
For each process, the slope
$b$ is given by the convolution of the transverse sizes of the
$q \bar q$ dipole, of the diffractively scattered system (which vanishes
for proton dissociation) and of the exchange (expected to be small).
As observed in Fig.~\ref{fig:b-slopes}, the elastic slopes for light VMs
strongly decrease with increasing \qsq, reflecting dipole shrinkage.
Values similar to those for \jpsi\ production are reached for 
$\scaleqsqplmsq\ \gapprox\ 5~\gevsq$.
A similar evolution is observed for DVCS as a function of \qsq.

The difference between elastic and proton dissociative slopes,
$b_{el} - b_{p. diss.}$, provides a test of proton vertex or ``Regge" factorisation.
It is of $3.5 \pm 0.1~\gevsqm$ for \jpsi, with a similar value for DVCS. 
The difference is higher, around $5.5$, for \rh\ and \ph\ mesons, with
however an indication of a decrease toward the \jpsi\ value 
with increasing $\scaleqsqplmsq$.
%%%%%%%%%%%%%%%%%%%%%%%%%%%%%
%%%%%%%%%%%%%%%%%%%%%%%%%%%%%
\paragraph {{\bf Energy dependence and effective Regge trajectory}}
%%%%%%%%%%%%%%%%%%%%%%%%%%%%%
The energy dependence of DVCS and VM production
is well described by a power law, $\rm {d} \sigma / \rm {d}W \propto W^{\delta}$.
For \jpsi\ photoproduction, the value of $\delta$ is significantly larger than 
for (soft) hadron--hadron interactions, with $\delta \sim 1.2$.
This confirms that \jpsi\ photoproduction is a hard process, characterised
by small transverse dipoles, which probes the low-$x$ gluon density in the 
proton at a scale where it is quickly increasing. 
For light VMs, the $W$ dependence is hardening with \qsq, 
and $\delta$ reaches \jpsi\ values for $\scaleqsqplmsq\ \gapprox\ 5~\gevsq$.

%--------------------------------------------------------------------------------------------------------
\begin{figure}[h]
\vspace{-0.5cm}
\begin{center}
\includegraphics[width=0.40\columnwidth]{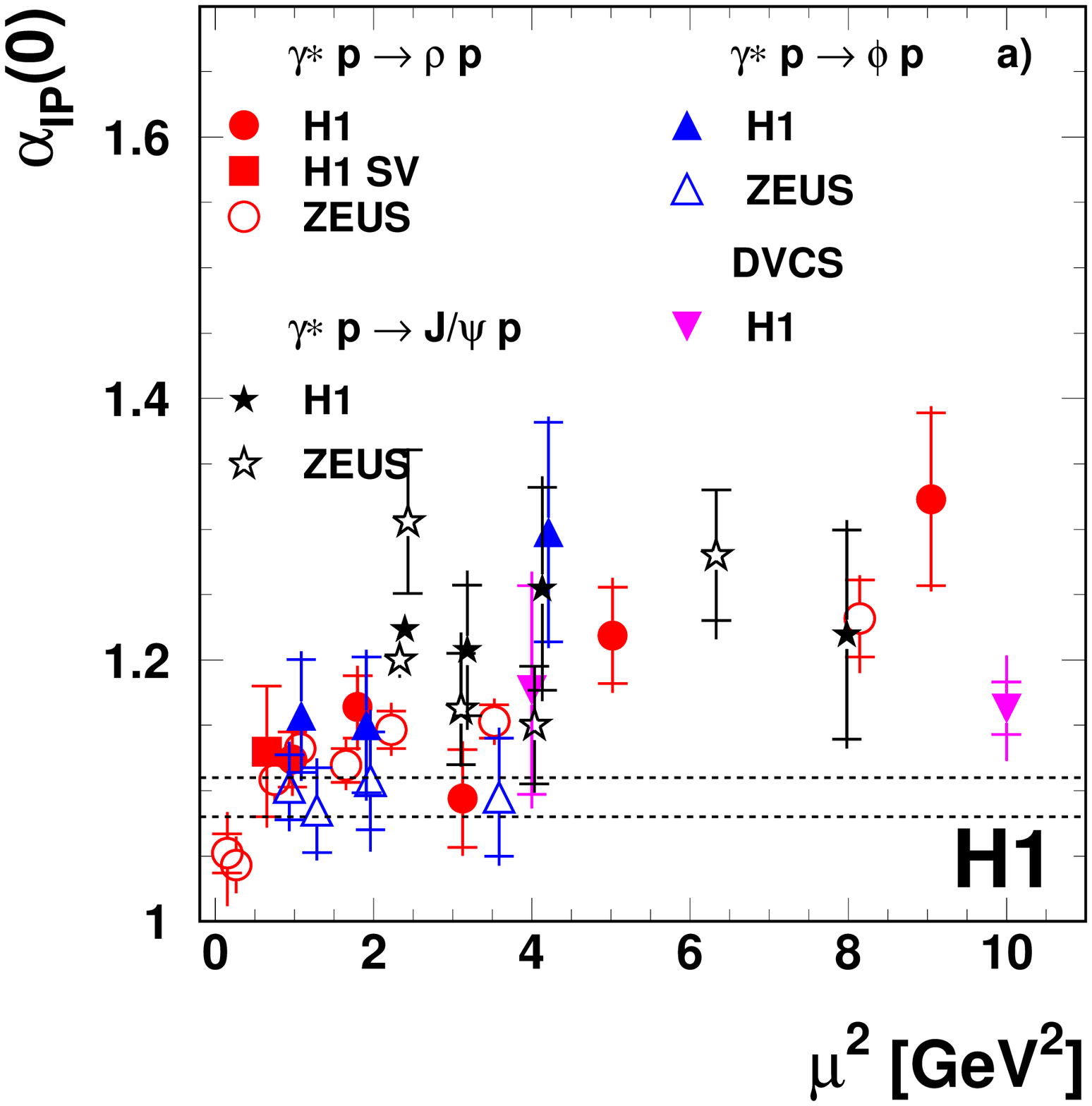}
\includegraphics[width=0.40\columnwidth]{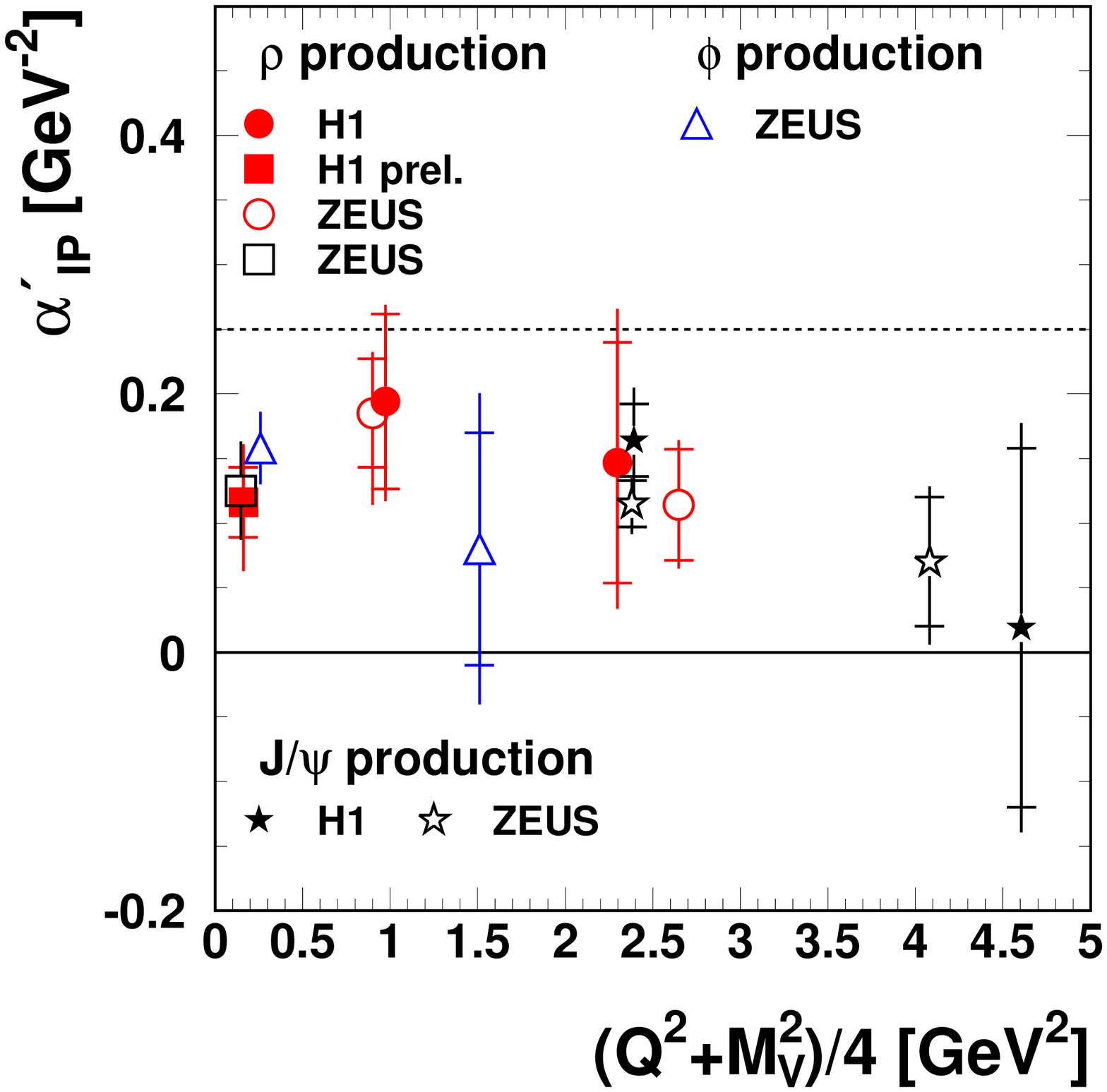}
\vspace{-0.3cm}
\caption{Measurement of (left) the intercept $\alpom(0)$ and (right) the slope \alp\
of the effective Regge trajectory, as a function of the scale 
$\mu^2 =  \qsqplmsq /4$ for VM production and $\mu^2 = \qsq$ for DVCS.}
\label{fig:Regge}
\end{center}
\vspace{-0.3cm}
\end{figure}
%--------------------------------------------------------------------------------------------------------
In a Regge inspired parameterisation, the energy dependence of the
cross section and its correlation with $t$ are given by
$\delta(t) =  4 \ ( \alpom(t)  - 1)$, with $\alpom(t) = \alpom(0) + \alp \cdot \ t$,
where \alp\ describes the shrinking of the diffractive peak with energy.
The hard behaviour of \jpsi\ production and the hardening with $\scaleqsqplmsq$ of 
the energy dependence of light VM production for $t = 0$ is demonstrated in 
Fig.~\ref{fig:Regge}, where values of $1.08$ or $1.11$ for $\alpom(0)$ are typical 
of soft hadron-hadron interactions.
The slope of the effective trajectory for VM production, including \rh\ 
photoproduction~\cite{rho-photoprod},
is smaller than the typical value $0.25~\gevsqm$ in hadronic interactions.
For DVCS $\alp = 0.03 \pm 0.09 \pm 0.11~\gevsqm$~\cite{dvcs}; 
for \jpsi\ photoproduction at high \modt, combining H1~\cite{high-t-jpsi} and ZEUS 
measurements, $\alp = -0.02 \pm 0.01 \pm 0.01~\gevsqm$.
%
%%%%%%%%%%%%%%%%%%%%%%%%%%%%%
%%%%%%%%%%%%%%%%%%%%%%%%%%%%%
\section {Helicity amplitudes}
%%%%%%%%%%%%%%%%%%%%%%%%%%%%%
%%%%%%%%%%%%%%%%%%%%%%%%%%%%%
The VM production and decay angular distributions allow the measurement of
fifteen spin density matrix elements, which are bilinear combinations of helicity
amplitudes.
Under natural parity exchange, five $T_{\lambda_V \lambda_{\gamma}}$ amplitudes 
are independent: two $s$-channel helicity conserving (SCHC) amplitudes 
($T_{00}$ and $T_{11}$), 
two single helicity flip amplitudes ($T_{01}$ and $T_{10}$) 
and one double flip amplitude ($ T_{-11}$).

\begin{wrapfigure}{l}{0.62\columnwidth}
\vspace{-0.5cm}
\begin{center}
\includegraphics[width=0.60\columnwidth]{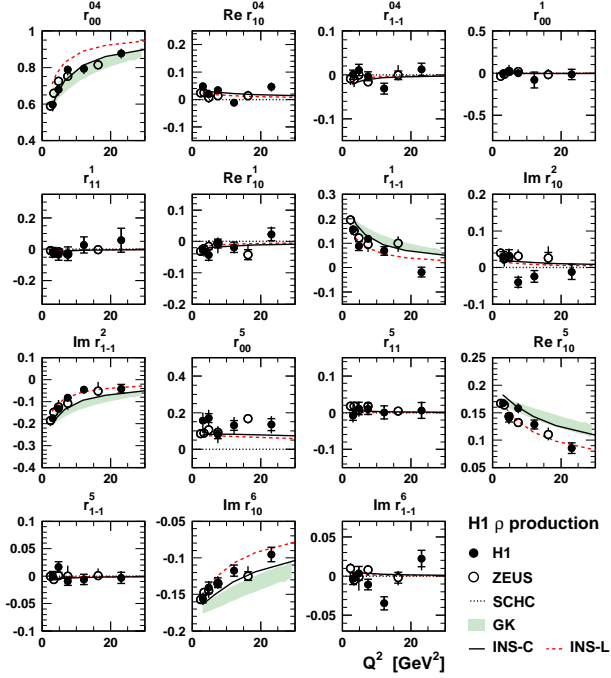}
\vspace{-0.3cm}
\caption{\qsq\ dependence of the spin density matrix elements for \rh\
electroproduction; model predictions are superimposed.
The dotted lines show the expected vanishing values 
if only the SCHC amplitudes are non-zero~\protect\cite{rho-hera1}.}
\label{fig:SDME}
\end{center}
\vspace{-0.5cm}
\end{wrapfigure}

The \qsq\ dependence of the matrix elements for \rh\ production is presented
in Fig.~\ref{fig:SDME}.
The five elements which contain products of the SCHC amplitudes 
are non-zero, whereas those formed with the helicity violating amplitudes are
generally consistent with 0.
A notable exception is the element \rczz, which
involves the product of the dominant $T_{00}$ SCHC 
%--------------------------------------------------------------------------------------------------------
%\begin{figure}[htbp]
%\end{figure}
%--------------------------------------------------------------------------------------------------------
amplitude with $T_{01}$,
which describes the transition from a transversely polarised photon to a longitudinal
\rh\ meson.

\begin{figure}[htbp]
%\vspace{-0.5cm}
\begin{center}
\includegraphics[width=0.32\columnwidth]{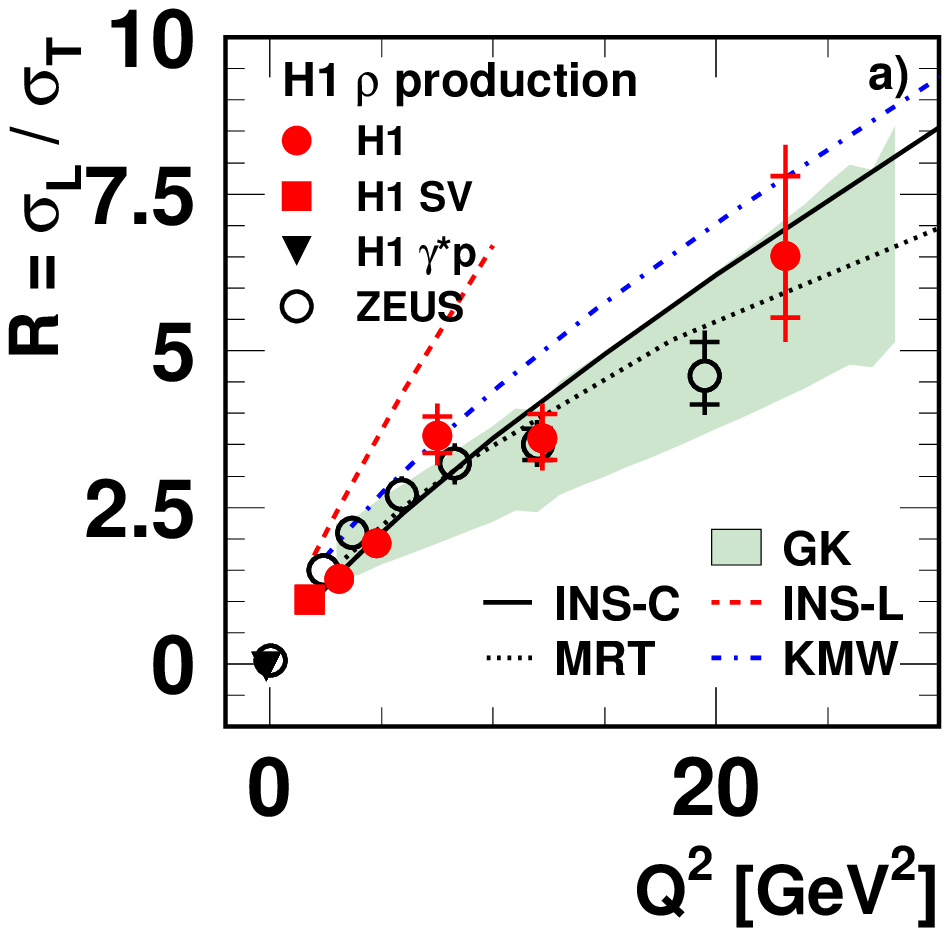}
\includegraphics[width=0.32\columnwidth]{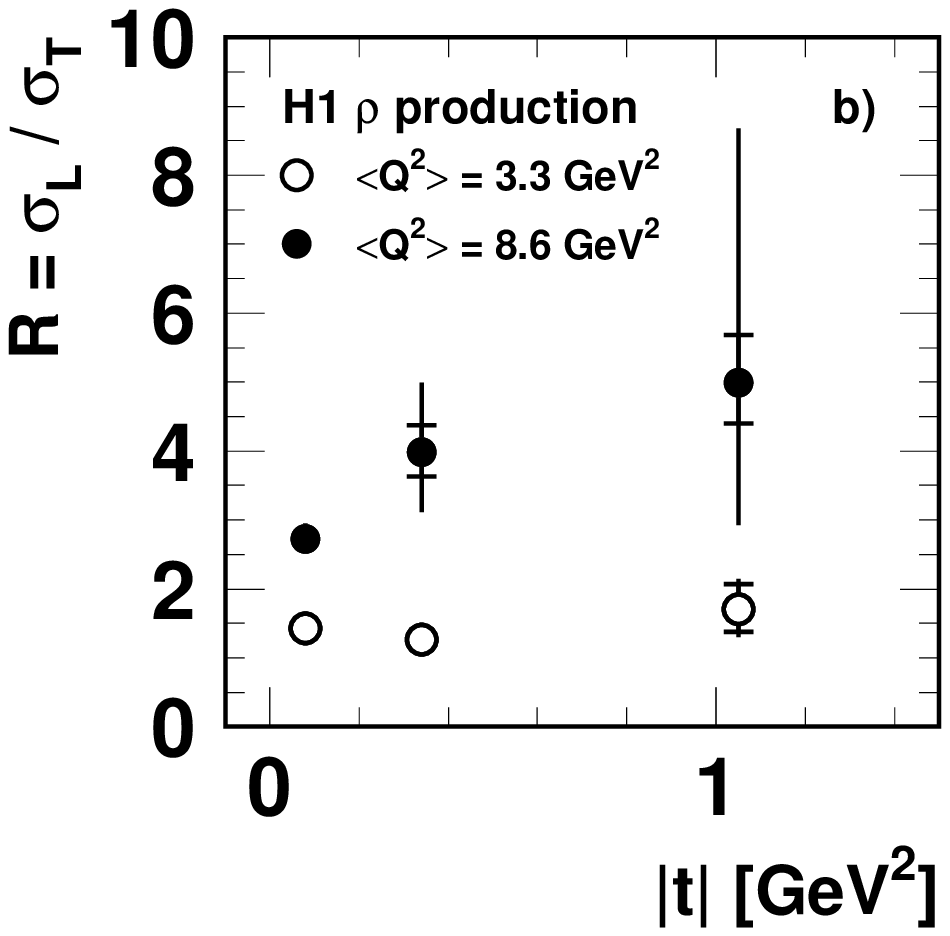}
\includegraphics[width=0.32\columnwidth]{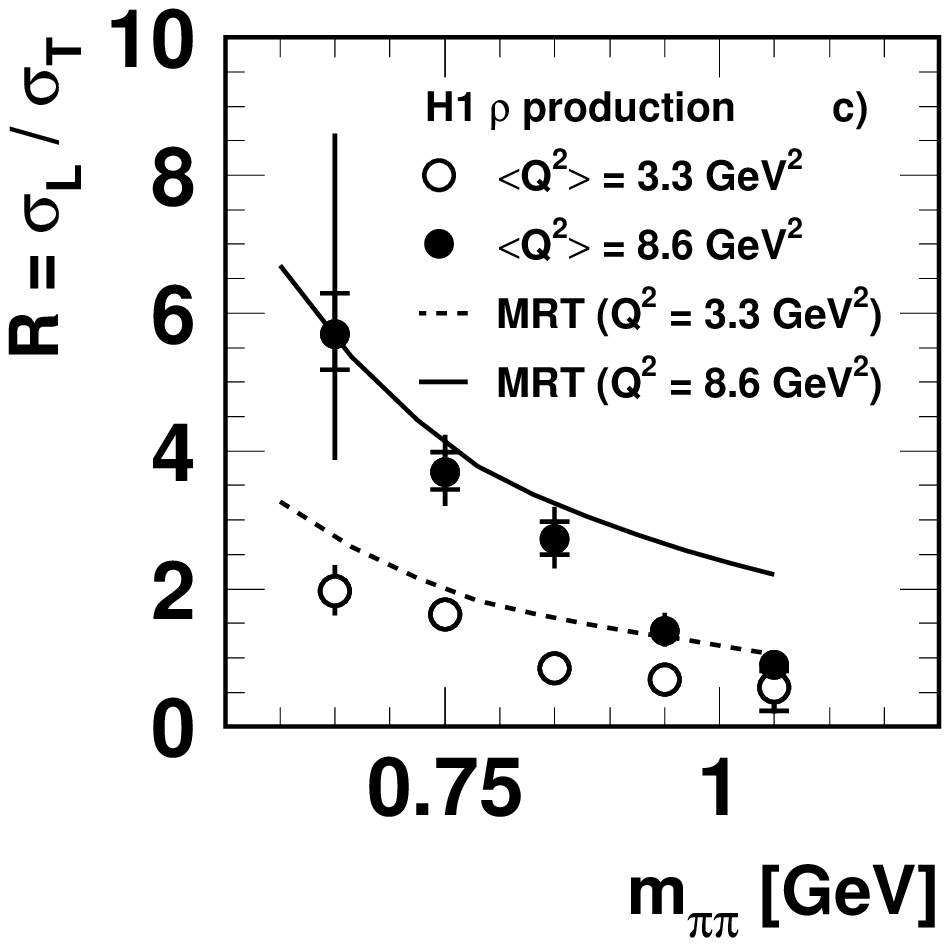}
\vspace{-0.3cm}
\caption{Measurement of $R = \sigma_L / \sigma_T$, as a function of \qsq, $t$ and 
the invariant mass of the two decay pions, for \rh\ 
electroproduction~\protect\cite{rho-hera1}; model predictions are superimposed.}
\label{fig:R}
\end{center}
\vspace{-0.3cm}
\end{figure}

The ratio $R = \sigma_L / \sigma_T$ of the longitudinal to transverse cross sections
for \rh\ production is shown is Fig.~\ref{fig:R}
as a function of \qsq\ and, for two domains in \qsq, as a function of $t$ and the 
invariant mass of the two decay pions.

A strong increase of $R$ with \qsq\ is observed, which is tamed at large 
\qsq. These features are relatively well described by models.
The \qsq\ dependence of $R$ for \rh, \ph\  and \jpsi\ VMs 
follows a universal trend when plotted as a function of 
$Q^2  / M_{V}^2$~\cite{rho-hera1}.

An increase of $R$ with $t$ is observed for $\qsq > 5~\gevsq$.
This can be translated into a measurement of the 
difference between the longitudinal and transverse $t$ slopes, through the relation
$R(t) = \sigma_L(t) / \sigma_T(t) \propto e^{- (b_L - b_T) |t|}$.
A slight indication ($1.5 \sigma$) is thus found for a negative value of $b_L - b_T$
($-0.65 \pm 0.14_{-0.51}^{+0.41}$) suggesting that, as expected, the
average transverse size of dipoles for transverse amplitudes is larger than for
longitudinal amplitudes

%--------------------------------------------------------------------------------------------------------
%--------------------------------------------------------------------------------------------------------
The dependence of $R$ with the dipion mass,  which can not 
be attributed solely to the interference of resonant \rh\ and non-resonant 
$\pi \pi$ production, 
indicates that the spin dynamics of \rh\ production depends of the effective
$q \bar q$ mass.    
No dependence of $R$ in $W$ is observed with the present data.

%--------------------------------------------------------------------------------------------------------
\begin{figure}[htbp]
\begin{center}
\includegraphics[width=0.19\columnwidth]{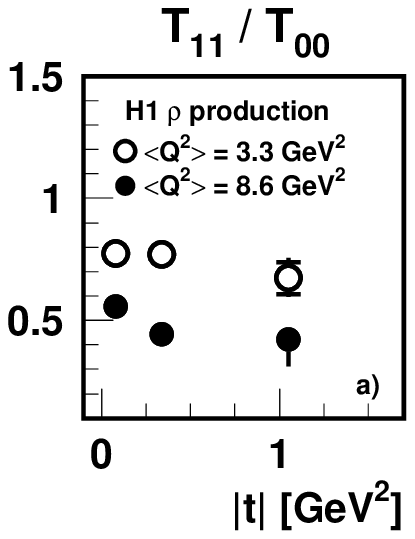}
\hspace {-0.3cm}
\includegraphics[width=0.19\columnwidth]{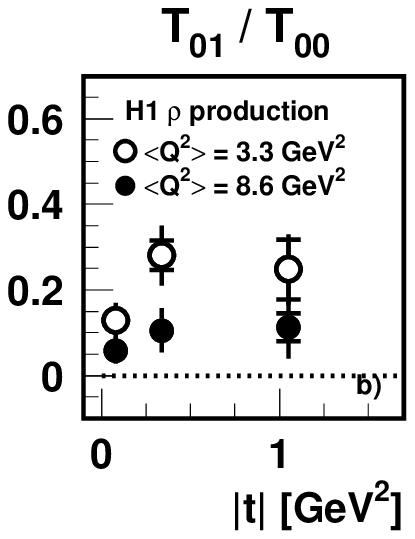}
\hspace {-0.3cm}
\includegraphics[width=0.19\columnwidth]{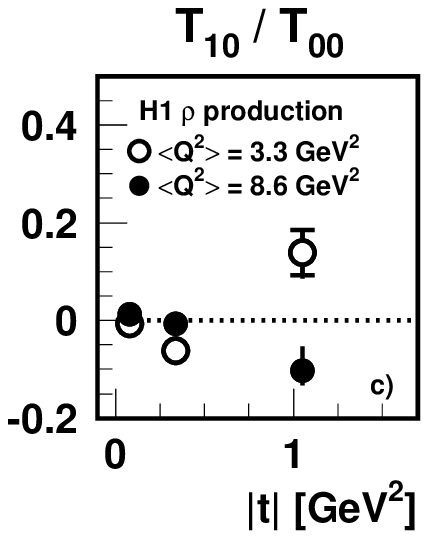}
\hspace {-0.3cm}
\includegraphics[width=0.19\columnwidth]{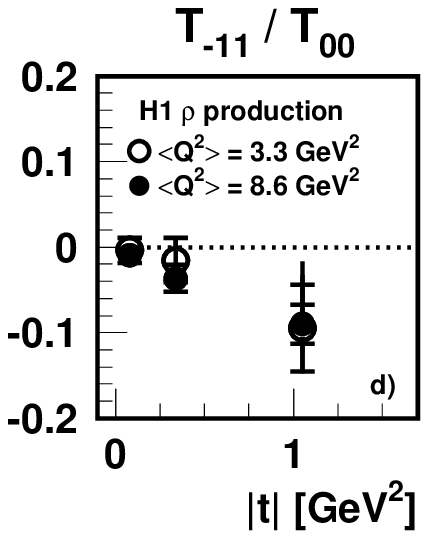}
\hspace {0.5cm}
\includegraphics[width=0.19\columnwidth]{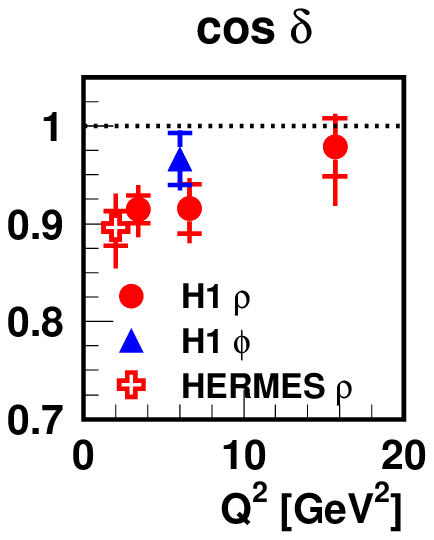}
\vspace{-0.3cm}
\caption{(a-d) Helicity amplitude ratios, as a function of $t$; (right plot)
phase difference between the two SCHC amplitudes, $T_{00}$ and 
$T_{11}$~\protect\cite{rho-hera1}.}
\label{fig:ampl_ratios_t}
\end{center}
\vspace{-0.5cm}
\end{figure}
%--------------------------------------------------------------------------------------------------------

Helicity amplitude ratios are measured, under the approximation that they are
in phase, through fits to the 15 matrix elements.
The four ratios to the dominant $T_{00}$ amplitude are presented in 
Fig.~\ref{fig:ampl_ratios_t} as a function of $t$, for two domains in \qsq.
At large \qsq, a $t$ dependence compatible with the expected $\sqrt{ \modt }$
law is observed for both single helicity flip amplitudes.
A significant double-flip amplitude $T_{-11}$ is observed, which may be related
to gluon polarisation in the proton.
The $t$ dependence of $T_{11}/T_{00}$ at large \qsq, a $3 \sigma$ effect, is related to the
$t$ dependence of $R$ and supports the indication of a difference between 
the transverse size of dipoles in transversely and longitudinally polarised photons.

A small non-zero phase difference between the two SCHC amplitudes, which decreases with
increasing \qsq, is visible in Fig.~\ref{fig:ampl_ratios_t}.
Through dispersion relations, this non-zero value is suggestive of different $W$ 
dependences of the longitudinal and transverse amplitudes.

\section*{Acknowledgments}
It is a pleasure to thank numerous colleagues from the ZEUS and H1 collaborations 
as well as theorists for enlightening discussions, 
and the workshop organisers and session conveners 
for the lively discussions and the pleasant atmosphere of the meeting.

\begin{footnotesize}

\end{footnotesize}

\end{document}